\documentclass[epj]{svjour}
\usepackage{graphics}
\usepackage{graphicx}%
\usepackage{multirow}%
\usepackage{amsmath,amssymb,amsfonts}%

\begin{document}

\title{Dispersion theory of nucleon (nucleus) Compton scattering spin polarizabilities and quasi-optical $\gamma$-ray polarization plane rotation and birefringence effect in a matter with polarized protons (nuclei)}

%\email{bar@inp.bsu.by}

\author{Vladimir Baryshevsky\inst{1} 
%	\thanks{\emph{Present address:} Insert the address here if needed}%
}                     % Do not remove
\offprints{}          % Insert a name or remove this line
\institute{Institute for Nuclear Problems, Belarusian State University, \\ 220030, 11 Bobruiskaya str., Minsk, Belarus, email:bar@inp.bsu.by,\,v\_baryshevsky@yahoo.com}
\date{Received: date / Revised version: date}

\abstract{
Experimental observation of
quasi-optical phenomenon of $\gamma$-quanta polarization plane rotation in matter
with polarized proton (nuclei) is demonstrated to be possible. This effect is similar to the
magneto-optic Faraday effect (Faraday optical rotation). Quasi-optical birefringence
effect for $\gamma$-quanta in matter with polarized nuclei having spin $S \ge 1$ could also be observed. The latter effect is similar to double refraction
(birefringence) of light in uniaxial and biaxial crystals. The above phenomena
open up possibility to investigate how spin vector and tensor polarizabilities
depend on $\gamma$-quanta energy. In particular, studying the effect of $\gamma$-quanta
polarization plane rotation, it is possible to find the range of $\gamma$-quanta energies,
which is associated with strong dependence of spin polarizability 
on $\gamma$-quanta energy.
%\PACS{
%	{PACS-key}{discribing text of that key}   \and
%	{PACS-key}{discribing text of that key}
%} % end of PACS codes
} %end of abstract

%\keywords{dispersion theory, vector nuclear spin polarizability, tensor spin polarizability, gamma ray plane rotation, gamma ray birefringence}



\maketitle

\section{Introduction}

Pioneering publications considering application of $\gamma$-quanta for investigation of the internal structure of a nucleon appeared in 50-th of the XXth century \cite{VG_1,57}.
%,VG_2}.
%
Those papers initiated numerous theoretical and experimental studies of this interesting possibility
 \cite{VG_8,VG_9,VG_9b,VG_3,VG_4,VG_5,VG_6,VG_7,VG_9c,VG_9d,VG_9e,VG_9f,VG_9i,VG_9_1,VG_n_10,VG_n_11,VG_n_12,VG_n_13,VG_n_14,VG_n_15,VG_n_16,VG_n_17,VG_n_18,VG_n_19,VG_n_21,VG_n_22,51,A1}.   
The first experiment on Compton scattering by the proton to measure the polarizabilities was carried out in 1958 (see \cite{VG_4,VG_5}).

Fast advancing of experimental methods  makes both single and double polarization experiments possible with a polarized solid target and polarized high-energy $\gamma$-quanta 
 \cite{VG_8,VG_9,VG_9b,VG_3,VG_4,VG_5,VG_6,VG_7,VG_9c,VG_9d,VG_9e,VG_9f,VG_9i,VG_9_1,VG_n_10,VG_n_11,VG_n_12,VG_n_13,VG_n_14,VG_n_15,VG_n_16,VG_n_17,VG_n_18,VG_n_19,VG_n_21,VG_n_22}. 
Polarized targets with high degree of nuclei polarization were created, namely: $NH_3$, $ND_3$ and $^6LiD$, as well as a butanol polarized proton (deuteron) target \cite{VG_n_13,VG_n_14,VG_n_15,VG_n_16,VG_n_17,VG_n_21,VG_n_22}.

In accordance with the pioneering results obtained by M.~Gell-Mann, M.L.~Goldberger and W.E.~Thirring \cite{57} the amplitude of forward elastic scattering  of a photon by a particle with spin $s=1/2$ can be written as follows:

\begin{equation}
	f_{\mu \nu}(\omega)=f_{1}(\omega)\left( \vec{e}^*_{\mu} \vec{e}_{\nu}  \right) + i
	f_{2}(\omega) \vec{\sigma}\left[ \vec{e}^*_{\mu} \vec{e}_{\nu}  \right],
	\label{eq:1}  
\end{equation}
where $\vec{e}_{\mu}$ and $\vec{e}_{\nu}$ are the photon polarization vectors in states $\mu$ and $\nu$, $\vec{\sigma}$ is the Pauli spin matrix.

Another notation is often used now when considering $\gamma$-quanta interaction with protons (nuclei), namely: $f_{1}(\omega)=f_{0}(\omega)$
and $f_{2}(\omega)=g_{0}(\omega)$. 
Therefore, for further consideration and analysis of (\ref{eq:1}) we will use it in the following form (see, for example \cite{VG_3,VG_4,VG_5,VG_6,VG_7})
\begin{equation}
	f_{\mu \nu}(\omega)=f_{0}(\omega)\left( \vec{e}^*_{\mu} \vec{e}_{\nu}  \right) + i
	g_{0}(\omega) \vec{\sigma}\left[ \vec{e}^*_{\mu} \vec{e}_{\nu}  \right].
	\label{eq:2}  
\end{equation}

According to analysis \cite{VG_3,VG_4,VG_5,VG_6,VG_7} equations (\ref{eq:1}) and (\ref{eq:2}) can be used to define electromagnetic polarizabilities and spin polarizabilities as the lowest order coefficients in an $\omega$-dependent development of the scattering amplitudes.
It was shown that the real part of amplitude $g_0 (\omega)$, which determines the spin-dependent part of amplitude $f_{\mu \nu}(\omega)$ at small frequencies can be expressed as:
\begin{equation}
	g_0 (\omega)= -\frac{e^2 \varkappa^2}{8\pi m^2} \omega +\gamma_0 \omega^3\, ,
	\label{eq:3}  
\end{equation}
where $\gamma_0$ is the spin polarizability, in the selected system of units electric charge $e$ meets the relation $\frac{e^2}{4 \pi}=\frac{1}{137.04}$, $\varkappa$ is the anomalous magnetic moment expressed in nuclear magneton. For a proton $\varkappa$=1.79, $\gamma_0 \sim -10^{-4}$ fm$^4$.
%\todo[inline]{see Table .....}

When frequency grows, expression (\ref{eq:3}) for $g_0(\omega)$, which includes constant $\gamma_0$ becomes invalid. 
Therefore, investigation of  $g_0(\omega)$ dependence on frequency $\omega$ for high-energy $\gamma$-quanta (energy range from few to dozens GeV or even higher) 
is of interest.

In present paper it is shown that
quasi-optical  phenomenon of $\gamma$-ray polarization plane rotation in a matter with spin polarized nuclei \cite{51,52,54,55,56,A2}  can be applied for $g_0(\omega)$ investigation along with experimental studies of scattering and different reactions excited by   $\gamma$-quanta, when they interact with polarized protons (nuclei).

Effect of polarization plane rotation for $\gamma$-quanta passing through matter with polarized spins was firstly theoretically described by Baryshevsky and Liuboshits in 1965 \cite{51}.
In details this effect was considered for $\gamma$-quanta passing through matter with polarized electrons.
Existence of this effect was confirmed by tests carried by Lobashev et al. \cite{54,55}, as well as by experimental studies of Bock and Luksch  \cite{56,57}. 
Energies of  $\gamma$-quanta in the above experiments  were as high as 230~keV, 290~keV and 330~keV.
It should be emphasized that the effect of polarization plane rotation is investigated in the beam moving in the direction of incident $\gamma$-quantum momentum, rather than scattered beam.
The effect is caused by $\gamma$-ray  refraction in matter.
It is interesting to spot that just as light in uniaxial and biaxial crystals
experiences double refraction (birefringence), so the similar quasi-optical
birefringence effect exists for $\gamma$-quanta in matter containing polarized nuclei
with spin $S \ge 1$.
This phenomenon enables studying tensor polarizability of a nucleus along with its spin  polarizability
% Это явление позволяет исследовать и тензорную поляризуемость ядра 
 \cite{A1,A2}.

The paper is organized as follows. 
%-------------------------
Section 2 demonstrates that phenomenon of $\gamma$-quanta polarization plane rotation in matter
	with polarized proton (nuclei) enables measuring the amplitude of elastic coherent forward scattering  and studying spin polarizability dependence on energy.
	Quasi-optical birefringence
	effect for $\gamma$-quanta in matter with polarized nuclei is shown in section 3 to enable measuring both spin and  tensor polarizabilities of nuclei.

%--------------from NO---------------------
%\bigskip

\section{Quasi-optical phenomenon of $\gamma$-quanta polarization plane rotation in matter with polarized proton (nuclei)}
\label{sec:rotation}

%\todo[inline]{-----------from NO}
%\section{\textcolor{red}{Gamma--optics of  polarized matter}}
%
%\subsection{\textcolor{red}{The Phenomenon of Rotation of the Polarization Plane of
%	$\gamma$-Quanta in Matter with Polarized Electrons. Magnetic X-ray
%	Scattering}}\label{sec:12}

It is well known that the phenomena caused by the optical anisotropy of matter (the Faraday effect, birefringence, natural
rotation of the light polarization plane) are eventually due to
the influence that the forces acting on electrons in atoms have on
the interaction of electromagnetic waves with matter.

Beyond the optical spectrum when the frequency of electromagnetic
waves becomes much greater than the average energy of electrons in
atoms and molecules, the interaction between radiation and matter
is reduced to the interaction of a photon with free electrons. As
a result, the structure of atoms and molecules becomes
non-essential, and hence the phenomena caused by optical
anisotropy of matter should disappear. For example, in the case of
the Faraday effect, a simple theory based on the normal Zeeman
effect gives the following expression for the light polarization
plane rotation angle $\vartheta$ per 1 cm path length \cite{4}:
\begin{equation}
	\label{12.1} \vartheta=\frac{\omega}{c}\frac{\partial
		n(\omega)}{\partial\omega}\frac{eB}{2mc}\,,
\end{equation}
where $n(\omega)$ - is the refractive index of matter in the
absence of a magnetic field; $e$ is the electron charge; $m$ is
its mass.

In a high--frequency range the expression for $n(\omega)$ can be written as follows (see \cite{4}):
\[
n (\omega)=1-\frac{2\pi e^{2}N}{m\omega^{2}}\,,
\]
where $N$ is the number of electrons per 1 cm$^3$ of matter, $m$ is the electron mass.
Thus we have for the rotation angle $\vartheta$ of a
high--energy quantum:
\begin{equation}
	\label{12.2} \vartheta=\frac{2\pi e^{3}N}{m^{2}c^{2}\omega^{2}}\,B\,.
\end{equation}
From (\ref{12.2}) we obtain that in the range of $\gamma$-quantum
energies of 100\,keV, the angle $\vartheta\approx 10^{-7}$\,rad/cm
for $B=10^{5}$ Gs=10 T, $ N\approx 10^{23}$ and decreases rapidly
with increasing quantum frequency.

However, in 1965, it was shown by  Baryshevsky and Lyuboshitz
\cite{51} that just at high energies of $\gamma$-quanta, there
becomes possible another mechanism of  photon  polarization plane
rotation in a target with polarized electrons, which is  due to
spin dependence of the amplitude of the elastic Compton forward
scattering of a $\gamma$-quantum by an electron. The calculations
showed that  the rotation angle reaches its maximum value of
$5.32\cdot10^{-3}$ rad/cm in the  range of quantum  energies from
500 to 700~ keV \cite{52}. This new phenomenon was experimentally
revealed by V.M. Lobashev, L.A. Popeko, L.M. Smotritskii, A.P.
Serebrov, E.A. Kolomenskii \cite{54,55} and  by Bock and Luksch
\cite{56}.

Let us consider the passage of a beam of  $\gamma$-quanta through
a medium with polarized electrons (nuclei). If the photon in right-- and
left--circularly polarized states has different  refractive
indices $n_{1}$ and $n_{2}$, then \cite{51,A2}:
\begin{equation}
	\label{12.3} \Delta n=n_2-n_1=\frac{2\pi
		Nc^{2}}{\omega^{2}}[f_{-}(0)-f_{+}(0)]\,,
\end{equation}
where $N$ is the number of electrons (nuclei) per unit volume; $f_{+}(0)$
and $f_{-}(0)$ are the spin--non-flip amplitudes of elastic
zero--angle scattering of right-- and left-- circularly polarized
photons by polarized electrons (nuclei), respectively;  $\omega$ is the
photon frequency.

The scattering amplitude for the Compton forward scattering by a particle with
spin $1/2$  can be written as follows \cite{57}:
\begin{equation}
	\label{12.4}
	f_{\mu\nu}=f_{1}(\omega)(\vec{e}\,^{*}_{\mu}\vec{e}_{\nu})+i
	f_{2}(\omega)\vec{\sigma}[\vec{e}\,^{*}_{\mu}\vec{e}_{\nu}]\,,
\end{equation}
where $\vec{e}_{\mu}$ and $\vec{e}_{\nu}$ are the photon
polarization vectors in states $\mu$ and $\nu$; $\vec{\sigma}/2$
is the particle spin operator. For the state with the right--hand
circular polarization
\begin{equation}
	\label{12.5} \vec{e}_{+}=-(\vec{e}_{1}+i \, \vec{e}_{2})/\sqrt{2}\,,
\end{equation}
while for the state with the left--hand circular polarization
\begin{equation}
	\label{12.6} \vec{e}_{-}=(\vec{e}_{1}-i \, \vec{e}_{2})/\sqrt{2}\,,
\end{equation}
Here $\vec{e}_{2}=[\vec{e}_{1}\vec{n}]$, where $\vec{n}$ is the
unit vector pointing in the propagation direction of the beam
of  $\gamma$-quanta.

In view of the above, it is easy to demonstrate
that
\begin{equation}
	\label{12.7} f_{+}=f_{1}(\omega) - f_{2}(\omega)(\vec{p}\vec{n}),
	f_{-}=f_{1}(\omega)+f_{2}(\omega)(\vec{p}\vec{n})\,,
\end{equation}
where $\vec{p}$ is the polarization vector of particles.

Suppose that photons in a vacuum are linearly polarized along the
direction  $\vec{e}_{1}$. Then in the  medium the polarization vector
$\vec{e}\,^{\,\prime}_{1}$ is
\begin{eqnarray}
	\label{12.8}
	\vec{e}\,^{\,\prime}_{1}&=&\left[\left(\frac{\vec{e}_{1}+i\vec{e}_{2}}{2}\right)\exp\left(-i\frac{\omega}{2c}\Delta n l \right)\right.\nonumber\\
	&+&\left.\left(\frac{\vec{e}_{1}-i\vec{e}_{2}}{2}\right)\exp\left(i\frac{\omega}{2c}\Delta n l \right)\right]\exp\left(i\frac{n_{1}+n_{2}}{2c}\omega l \right)\nonumber\\
	&=&\exp\left(i{\omega}\frac{n_{1}+n_{2}}{c} l \right)\left[\vec{e}_{1}\cos\left(\frac{2\pi Nc}{\omega}(\vec{p}\vec{n})f_{2}(\omega) l \right)\right.\nonumber\\
	&+&\left.\vec{e}_{2}\sin\left(\frac{2\pi
		Nc}{\omega}(\vec{p}\vec{n})f_{2}(\omega) l \right)\right]\,,
\end{eqnarray}
where $l$ is the 
path passed by $\gamma$-quanta in matter.
%distance from the boundary between vacuum and the electron target,
%which is counted off along the direction of photon propagation
%in the medium.

%If $\texttt{Im}f_{2}(\omega)=0$, which is only possible at a
%certain value of frequency $\omega$, a pure rotation of the
%polarization plane of photons occurs. 

Full rotation of the
polarization vector takes place in the length
\begin{equation}
	\label{12.9} d=\frac{4\pi c}{\omega|\Delta n|}\,.
\end{equation}
One can easily see  that the positive sign of
$f_{2}(\omega)(\vec{p}\vec{n})$ corresponds to the right--hand
rotation, while the negative sign of this quantity corresponds to the left--hand rotation.

In the general case, $\texttt{Im}f_{2}(\omega)\neq 0$, i.e., the
coefficients of absorption are different for the states with left and
right circular polarizations. As in this case
(\ref{12.8}) includes the trigonometric functions for complex
arguments, the dependence of photon polarization on  distance
$l$ becomes more complicated. If the photons in a vacuum  are still
polarized along the direction  $\vec{e}_1$, the following formulas are
applicable to the Stokes parameters in a medium \cite{3,23}
\begin{equation}
	\label{12.10} \varepsilon_{1}=r\cos 2\varphi;~~
	\varepsilon_{2}=(1-r^{2})^{1/2};~~\varepsilon_{3}= r \sin 2\varphi\,,
\end{equation}
where
\begin{eqnarray*}
	\varphi & = &\frac{2\pi N
		c}{\omega}(\vec{p}\vec{n})\texttt{Re}f_{2}(\omega)\,l\,;\\
	r & = &\cosh\left(\frac{4\pi N
		c}{\omega}(\vec{p}\vec{n})\texttt{Im}f_{2}(\omega)\, l \right)=(\varepsilon_{1}^{2}+\varepsilon_{3}^{2})^{1/2}
\end{eqnarray*}
is the degree of linear polarization; $\varepsilon_{2}$ is the
degree of circular polarization. At $l=0$, we have  $\varepsilon_{1}=1$,
$\varepsilon_{3}=0$, $\varepsilon_{2}=0$.

It is seen that when the imaginary part of the function
$f_{2}(\omega)$ is nonzero, the  linear polarization of the  photon in a
medium transforms into an elliptical one, and $\varphi$ is the angle of rotation
of the ellipse's major axis  relative to the initial direction
$\vec{e}_{1}$.

It follows from the above that
the full rotation of the ellipse's major axis occurs in the length
\begin{equation}
	\label{12.11} d=\left(\frac{N
		c}{\omega}(\vec{p}\vec{n})\,\texttt{Re}f_{2}(\omega)\right)^{-1}\,.
\end{equation}
Note that at $l\rightarrow\infty$, we have $|\varepsilon_{2}|=1$.
This indicates the total absorption of photons with right-- or
left--hand circular polarization.

It immediately follows from  (\ref{12.10}) that the
change in the polarization of $\gamma$-quanta  passing through a
polarized target only depends on the function
$f_{2}(\omega)$. As for the function $f_{1}(\omega)$ [see
(\ref{12.4})], it has nothing to deal with the effect we are
concerned with.

%----- end of part 1 ---

From the
optical theorem follows the below relation
\begin{equation}
	\label{12.12} \texttt{Im}f_{2}(\omega)=\frac{\omega}{4\pi
		c}\frac{\sigma_{\uparrow\downarrow}(\omega)-\sigma_{\uparrow\uparrow}(\omega)}{2}\,,
\end{equation}
where $\sigma\uparrow\uparrow$ and $\sigma\uparrow\downarrow$ are
the values of the total Compton scattering  cross sections for
parallel and antiparallel orientations of photon and electron (nucleus)
spins, respectively. 

%------ end of part 2

To calculate the real part of $f_{2}(\omega)$, make use of the
dispersion relation given in \cite{57}.
\begin{equation}
	\label{12.14} \texttt{Re}
	f_{2}(\omega)=-\frac{2\omega}{\hbar c^{2}}(\Delta\mu)^{2}+\frac{2\omega^{3}}{\pi}\int\limits_{0}^{\infty}\frac{\texttt{Im}f_{2}(\omega^{\,\prime})}{\omega^{\,\prime
			2}(\omega^{\,\prime 2}-\omega^{2})}d\omega^{\,\prime}\,,
\end{equation}
where $\Delta\mu$  is the anomalous magnetic moment of the particle.

%---- end of part 3
%\todo[inline]{end from NO}

Using expression (\ref{12.12}) for $\texttt{Im}f_{2}(\omega)$ one can express $\texttt{Re}f_{2}(\omega)$ as follows:
\begin{equation}
	\label{eq:4}
	\texttt{Re}
	f_{2}(\omega)=-\frac{2 \omega}{\hbar c^{2}}(\Delta\mu)^{2}+\frac{\omega^{3}}{4\pi^{2}c}\int\limits_{0}^{\infty}\frac{\sigma_{\uparrow\downarrow}(\omega^{\,\prime})
		-\sigma_{\uparrow\downarrow}(\omega^{\,\prime})}{\omega^{\,\prime}(\omega^{\,\prime
			2}-\omega^{2})}d\omega^{\,\prime}\,.
\end{equation}
According to expression (\ref{12.10}) the polarization plane rotation angle for $\gamma$-quanta moving through a polarized target is determined by $\texttt{Re}
f_{2}(\omega)$. Expression (\ref{eq:4}) for $\texttt{Re}
f_{2}(\omega)$ comprises two summands both conditioned by scattering: the first one is caused by anomalous magnetic moment $\Delta\mu$, while the second is associated with other scattering processes and reactions caused by $\gamma$-quanta interactions with protons (nuclei, electrons).

In accordance with expressions (\ref{12.10}) and (\ref{eq:4})
polarization plane rotation angle $\varphi$ for a $\gamma$-quantum, which passed in matter   path $l$, reads as follows:
\begin{equation}
	\label{eq:5}
\varphi=-\frac{4 \pi N}{\hbar c}(\Delta \mu)^2 (\vec{p} \vec{n}) l + \frac{2 \pi N c}{\omega} (\vec{p} \vec{n}) l \frac{2 \omega^3}{\pi} \int_{0}^{\infty} \frac{\texttt{Im} f_2(\omega^{\,\prime})}{\omega^{\,\prime \,2} (\omega^{\,\prime \, 2}-\omega^{ 2})} d \omega^{\,\prime}
\,.
\end{equation}
Expression (\ref{eq:5}) can be rewritten as follows:
\begin{equation}
	\label{eq:6}
	\varphi=-\frac{4 \pi N}{\hbar c}(\Delta \mu)^2 (\vec{p} \vec{n}) l + {2 \pi N c} (\vec{p} \vec{n}) l \gamma_0 (\omega) \omega^{2}
	\,,
\end{equation}
where 
$$\gamma_0 (\omega)=\frac{1}{4 \pi^2 }  \int_{0}^{\infty} \frac{\sigma_{\downarrow \uparrow}(\omega^{\,\prime}) - \sigma_{\uparrow \uparrow}(\omega^{\,\prime})}{\omega^{\,\prime} (\omega^{\,\prime \,2}-\omega^{ 2})} d \omega^{\,\prime}.
$$

In the range of low energies according to 
 \cite{VG_4,VG_5,VG_6,VG_7,VG_9c,VG_9d,VG_9e,VG_9f,VG_9i,VG_9_1,VG_n_10,VG_n_11,VG_n_12,VG_n_13,VG_n_14,VG_n_15,VG_n_16,VG_n_17,VG_n_18,VG_n_19}
the amplitude of Compton forward scattering can be expressed in the form (\ref{eq:3}) with $\gamma_0$ read as follows:
\begin{equation}
	\label{eq:8}
\gamma_0 (\omega)=\frac{1}{4 \pi^2 }  \int_{\omega_{thr}}^{\infty} \frac{\sigma_{\downarrow \uparrow}(\omega^{\,\prime}) - \sigma_{\uparrow \uparrow}(\omega^{\,\prime})}{\omega^{\,\prime \,3} } d \omega^{\,\prime},
\end{equation}
where $\omega_{thr}$ is the pion photoproduction threshold. 
Contributions to the total cross-sections, caused by Compton scattering  and processes of electron-positron pair production  are not considered in (\ref{eq:8}).

Carried out experiments and theoretical analysis provide the following evaluation for proton spin polarizability $\gamma_0$:
$\gamma_0 \approx -1.34 \cdot 10^{-4}$\,fm$^4$\,$=1.34 \cdot 10^{-56}$\,cm$^4$.
Therefore, in case if (\ref{eq:3}) is valid, polarization plane rotation angle $\varphi$ for a $\gamma$-quantum, which passed in polarized matter   path $l$, reads as follows:
\begin{equation}
	\label{eq:9}
	\varphi=-\frac{4 \pi N}{\hbar c}(\Delta \mu)^2 (\vec{p} \vec{n}) l + {2 \pi N} (\vec{p} \vec{n}) \gamma_0 k^2 l =\varphi(\Delta \mu)+\varphi(\gamma_0)
	\,,
\end{equation}
where $k=\frac{\omega}{c}$ is the wavenumber of the $\gamma$-quantum.

The summand $\varphi(\Delta \mu)$, which includes anomalous magnetic moment, does not depend on energy, while  another one $\varphi(\gamma_0)$, which is determined by  spin polarizability $\gamma_0$,
depends on $\gamma$-quantum energy.
The latter grows proportionally to $k^2$ i.e. proportionally to the squared $\gamma$-quantum energy.

Let us now evaluate the value of polarization plane rotation angle $\varphi$ for a $\gamma$-quantum, which passes through a target with polarized nuclei.
Target $^{14}$NH$_3$, which is used for investigation of polarized $\gamma$-quanta  scattering  by polarized protons, can be considered as an example to evaluate rotation effect.
Note, that this target comprises both protons and nuclei $^{14}$N,
therefore, number of nuclei per cm$^3$ in this target is not equal to number of protons.
To evaluate the number of polarized protons per cm$^3$ let us use the density $\rho$ in units $[\frac{g}{cm^3}]$, which for solid NH$_3$ is as high as $\rho=0.85 [\frac{g}{cm^3}]$.
The weight of  NH$_3$ molecule is with high accuracy equal to $M=17m_p$, where $m_p$ is the proton mass. Therefore, the number of NH$_3$ molecules per cm$^3$ reads $N_{mol}=\frac{\rho}{17m_p}$. And, since each molecule comprises three protons, the number of protons can be expressed as 
$$
N=3 \cdot N_{mol}= \frac{\rho}{m_p} \cdot \frac{3}{17}= \frac{\rho}{m_p} \cdot f.
$$ 
Here $f$ is the dilution factor (see, for example, \cite{VG_n_13,VG_n_14,VG_n_21,VG_n_22}) and $\frac{\rho}{m_p}$ is the number of nucleons per cm$^3$.
The complicated internal structure of the target requires the target density to be reduced by the so called \cite{VG_n_13,VG_n_14,VG_n_21,VG_n_22} packing factor $\varkappa=0.6$ resulting in correction in the number of protons
per cm$^3$
as follows:
$$
N=\frac{\rho}{m_p} \cdot f \cdot \varkappa.
$$
Therefore, 
polarization plane rotation angle $\varphi$ is finally expressed as follows:
\begin{equation}
	\label{eq:10}
	\varphi=-\frac{4 \pi }{\hbar c}\frac{\rho}{m_p} f \varkappa p (\vec{n}_p \vec{n}) (\Delta \mu)^2  l + \frac{2 \pi \rho }{m_p} f \varkappa p (\vec{n}_p \vec{n})  \gamma_0 k^2 l 
	\,,
\end{equation}
where $p$ is the proton polarization degree, $\vec{n}_p$ is the unit vector directed along proton polarization vector.
%
%It should be mentioned that product $fp$ is called the effective  polarization for protons and is denoted as \textcolor{red}{$P_{eff}$} \cite{bibid}.
%\todo{check notation}

If $\gamma$-quanta momentum is directed along the polarization vector, then $\vec{n}_p \vec{n}=+1$, in case of antiparallel directed $\vec{n}_p$ and $\vec{n}$, product $\vec{n}_p \vec{n}=-1$.
Therefore, change of $\vec{n}_p$ direction with respect to $\vec{n}$ results in change of rotation direction (sign).
For $\vec{n}_p \uparrow \uparrow \vec{n}$ polarization plane rotation angle reads as
\begin{equation}
	\label{eq:11}
	\varphi=-\frac{4 \pi \rho \varkappa}{\hbar c \,m_p} f  p \, (\Delta \mu)^2  l + \frac{2 \pi \rho \varkappa}{m_p} f  p \, \gamma_0 k^2 l
%	 = 	\varphi(\Delta \mu) + \varphi(\gamma_0)
	\,.
\end{equation}
Let us evaluate rotation angle $\varphi$.
In case of NH$_3$ target with polarized protons the number of protons per cm$^3$ is $N=\frac{\rho \varkappa f}{m_p} \approx 5.4 \cdot 10^{22}$, therefore at  polarization degree $p=0.9$  the number of polarized protons in the target is as high as $N_p \approx 5 \cdot 10^{22}$.
Anomalous magnetic moment $\Delta \mu = 8.95 \cdot 10^{-24}$\,erg/Gs, therefore, contribution to rotation angle $\varphi(\Delta \mu)$ can be evaluated as 
\begin{equation}
	\label{eq:12}
\varphi(\Delta \mu)=1.6 \cdot 10^{-6} \cdot l ~\textrm{rad}
\end{equation}
and for $l=$30~cm the angle $\varphi(\Delta \mu) \approx 5 \cdot 10^{-5}$\,rad. 
Contribution to rotation angle $\varphi(\Delta \mu)$ caused by anomalous magnetic moment does not depend on $\gamma$-quantum energy.

Let us now evaluate contribution to rotation angle $\varphi(\gamma_0)$, which is determined by  spin polarizability $\gamma_0$ and
depends on $\gamma$-quantum energy.
Using the second summand in (\ref{eq:11}) one can get for $\varphi(\gamma_0)$ in case of NH$_3$ target the following evaluation:
\begin{equation}
	\label{eq:13}
	\varphi(\gamma_0) \approx 3 \cdot 10^{-33} k^2\, l.
\end{equation}
It follows from (\ref{eq:13}) that for $\gamma$-quanta with energy 300~MeV ($k=1.6 \cdot 10^{13}$\,cm$^{-1}$) passing through the target of 30~cm thickness rotation angle $\varphi(\gamma_0)$ is as high as $\varphi(\gamma_0)=2 \cdot 10^{-5}$\,rad.
If $\gamma$-quanta energy is increased up to 1~GeV (such energies are available at the Bonn accelerator facility ELSA and at the Mainz accelerator MAMI
%24,25,32,33
\cite{VG_n_13,VG_n_14,VG_n_21,VG_n_22}) 
this contribution could reach $\varphi(\gamma_0)=2 \cdot 10^{-4}$\,rad. 
Further increase of $\gamma$-quanta energies results in fast growth of rotation angle: for 3~GeV $\gamma$-quanta $\varphi(\gamma_0)=2 \cdot 10^{-3}$\,rad and for 100~GeV it appears to be $\varphi(\gamma_0)=1.8$\,rad  for the target of 30~cm thickness.

The above evaluations for angle of $\gamma$-quanta polarization plane rotation at the GeV-scale energies available at the facilities, where investigation of Compton interaction of polarized $\gamma$-quanta with polarized protons   \cite{VG_n_13,VG_n_14,VG_n_21,VG_n_22} are carried out, therefore, the effect of $\gamma$-quanta polarization plane rotation can be experimentally observed.
Thus, law (\ref{eq:3})  can be verified and the range, for which dependence of amplitude $\gamma_0(\omega)$ on $\gamma$-quanta energy should be taken into account, can be defined.

Let us emphasize that investigation of quasi-optical effect of $\gamma$-quanta polarization plane rotation does not imply measurement of properties of scattered $\gamma$-quanta (scattered electromagnetic wave).
The considered effect is studied by investigation of coherent passing of $\gamma$-quanta (electromagnetic wave) through a target  in the direction of the initial momentum of $\gamma$-quanta absolutely similar to the magneto-optic Faraday effect (Faraday optical rotation).
The number of $\gamma$-quanta passed through the target appears to be much higher as compared with the number of those scattered into some spatial angle.

\section{Quasi-optical birefringence effect for $\gamma$-quanta in matter with polarized nuclei}
\label{sec:birefringence}

It is interesting to spot that just as light in uniaxial and biaxial crystals experiences double refraction (birefringence), so the similar quasi-optical birefringence effect exists for  $\gamma$-quanta in matter containing polarized nuclei with spin {$S \ge 1$} \cite{A1,A2}. 

For example, for $^{14}$N and D spin is equal $S=1$, while $^7$Li has $S=3/2$.

According to the {above} 
%
%\todo{see, for example, (\ref{12.3})}
%
the index of $\gamma$-quanta in refraction in matter is determined by the coherent elastic forward scattering amplitude $f(0)$ as follows:
\begin{equation*}
	n=1+\frac{2 \pi \rho}{k^2} {f(0)}.
\end{equation*}

When $\gamma$-quanta interact with matter, which comprises polarized nulei with 
spin {$S \ge 1$},
amplitude $f(0)$ can be expressed as {\cite{A1,A2}}:
\begin{eqnarray}
	\label{eq:14}
	f(0)=f_1(\omega) (\vec{e}^{\,\,\prime*}~\vec{e}) + i \, f_2(\omega) \, \vec{p} \, \left[\vec{e}^{\,\,\prime*}~\vec{e}\right] + \nonumber \\ {f_3 (\omega)} \, Q_{ik} {e}^{\,\,\prime*}_i  e_k + f_4(\omega) n_{\gamma i} n_{\gamma k} Q_{ik}.
\end{eqnarray}
where $\vec{p}=\textrm{Tr} \hat{\rho} \vec{n}_S$ is the nuclear polarization vector, $\vec{n}_S= \hat{\vec{S}}/S$, $\hat{\vec{S}}$ is the nuclear spin operator, $\hat{\rho}$ is the spin density matrix of the target, $Q_{ik}=\textrm{Tr} \hat{\rho}\, \hat{Q}_{ik}$ is the polarization tensor of rank two, 
$$
\hat{Q}_{ik}=\frac{3}{2 S (2S-1)}\left\{  \hat{S}_i \hat{S}_k + \hat{S}_k \hat{S}_i - \frac{2}{3}S (S+1) \delta_{ik}     \right\} ;
$$
$\vec{n}_{\gamma}$ is the unit vector along the $\gamma$-quanta momentum.

Let us assume that polarization vector $\vec{p}$ for a target is orthogonal to $\gamma$-quanta incidence direction: $\vec{p} \perp \vec{n}_{\gamma}$\,, and define direction of $\vec{n}_{\gamma}$  as axis $y$.
In this case from (\ref{eq:14}) one could observe the difference in refraction indices for a photon with linear polarization $\vec{e}_x$ and that with linear polarization $\vec{e}_z$.

Let linear polarization of a $\gamma$-quantum incident on a target is a superposition $\vec{e}= \alpha \vec{e}_x + \beta \vec{e}_z$\,.
As the $\gamma$-quantum moves deeper into target, its linear polarization  converts into  elliptical one that is in full similarity to optics.
Thus, circular polarization appears in the initially linearly polarized  beam of $\gamma$-quanta. Polarization degree is determined by $\textrm{Re}\,f_3(\omega)$ and can be expressed via tensor polarizability of the nucleus, which was introduced in {\cite{VG_8,VG_9}}.

In case when a $\gamma$-quantum with circular polarization moves in the target,  it attains linear polarization, degree of which is determined by $\textrm{Re}\,f_3(\omega)$.
The detailed description of birefringence for $\gamma$-quanta is given in \cite{A1,A2}.

According to evaluations \cite{A1} in the vicinity of  giant resonance, the degree $\delta$ of attained circular polarization for a $\gamma$-quantum, which initially has linear polarization, when it passes through a target with $l=1$~cm thickness comprising $\textrm{Ta}$ nuclei, is as high as $\delta \approx 10^{-3} - 10^{-4}$.
For a target comprising polarized deuteron nuclei polarization degree could be evaluated as $\delta \approx 4 \cdot10^{-5}\, l$,
assuming that $\gamma$-quanta energy is about few MeV and deuteron tensor polarizability is evaluated by its  static polarizability $\alpha_T \approx 10^{-40}$\,cm$^3$ derived in {\cite{A3}}.

For other nuclei and $\gamma$-quanta energies as high as dozens and hundreds of MeV and higher, the birefringence effect should be evaluated separately.

\section{Conclusion}
\label{sec:conclusion}
Success in development of targets with polarized nuclei and beams of polarized high-energy $\gamma$-quanta  enables direct experimental observation of  quasi-optical phenomenon of $\gamma$-quanta polarization plane rotation in matter with polarized proton (nuclei).
This effect is similar to the magneto-optic Faraday effect (Faraday optical rotation).
Quasi-optical birefringence effect  for $\gamma$-quanta in matter with polarized nuclei having spin $S \ge 1$ could also be observed. The latter effect is similar to double refraction (birefringence) of light in uniaxial and biaxial crystals.
The above phenomena open up possibility to investigate how spin vector and tensor polarizabilities depend on $\gamma$-quanta energy.
In particular, studying the effect of $\gamma$-quanta polarization plane rotation, it is possible to find the range of  $\gamma$-quanta energies, which is associated with strong dependence  of spin polarizability $\gamma_{0}$ on $\gamma$-quanta energy. For this energy range fast growth of the amplitude of forward scattering $g_0(\omega) \sim \gamma_0 \, \omega^3$ with $\gamma$-quanta energy growth is expected to be slowed.

%----------------------------------------------------

\end{document}